\documentclass[sigconf]{acmart}
\AtBeginDocument{%
  }

\setcopyright{acmlicensed}
\copyrightyear{2018}
\acmYear{2018}
\acmDOI{XXXXXXX.XXXXXXX}
\acmConference[Conference acronym 'XX]{Make sure to enter the correct
  conference title from your rights confirmation email}{June 03--05,
  2018}{Woodstock, NY}
\acmISBN{978-1-4503-XXXX-X/2018/06}

\usepackage{comment}
\usepackage{url}
\usepackage{booktabs}
\usepackage{subcaption}
\usepackage{multirow} 
\usepackage[most]{tcolorbox}
\usepackage{footmisc}
\usepackage[dvipsnames]{xcolor}

\newtcolorbox{insightbox}[1]{
    colback=orange!10,      
    colframe=orange!70, 
    colbacktitle=orange!70, 
    title={#1},             
    fonttitle=\sffamily\bfseries\large,
    fontupper=\sffamily,    
    arc=15pt,               
    outer arc=15pt,
    left=15pt,              
    right=15pt,
    top=10pt,
    bottom=10pt,
    boxrule=1.5pt,          
    titlerule=0pt,          
    toptitle=5pt,
    bottomtitle=2pt,
    enhanced,               
}
\begin{document}


\title{From Notepad AI to Social Media: How Can Text Style Transformation Mitigate Social Harm?}

\author{Syed Mhamudul Hasan}
\affiliation{%
  \institution{Southern Illinois University}
  \city{Carbondale}
  \state{IL}
  \postcode{62901}
  \country{USA}}
\email{syedmhamudul.hasan@siu.edu}

\author{Mohd. Farhan Israk Soumik}
\affiliation{%
  \institution{Southern Illinois University}
  \city{Carbondale}
  \state{IL}
  \postcode{62901}
  \country{USA}}
\email{mohdfarhanisrak.soumik@siu.edu}

\author{Abdur R. Shahid}
\affiliation{%
  \institution{Southern Illinois University}
  \city{Carbondale}
  \state{IL}
  \postcode{62901}
  \country{USA}}
\email{shahid@cs.siu.edu}

\renewcommand{\shortauthors}{Hasan et al.}



\begin{abstract}
The rapid proliferation of harmful and emotionally damaging content on social media platforms has intensified concerns regarding societal harm. While content moderation efforts primarily focus on detecting and removing harmful posts, less attention has been given to mitigating harm through stylistic text transformation while preserving semantic meaning. In this paper, we propose a writing-assistance framework that can reduce societal harm by transforming aggressive, toxic, or emotionally harmful comments into softer, more neutral stylistic forms inspired by \textbf{Notepad AI}, a simple AI writing assistant. Rather than censoring or suppressing speech, we apply controlled stylistic modifications to preserve core informational content while reducing emotional intensity and identity-based attacks. We introduce an \textbf{Emotion Drift Index (EDI)} metric to systematically quantify emotional change and evaluate the effectiveness of stylistic rewriting, thereby reducing harmful interactions in online environments.
\end{abstract}

\ccsdesc[500]{Computing Methodologies~Natural Language Generation}
\ccsdesc[300]{Information Systems~Social Networks}
\ccsdesc[300]{Social Harm}
%

\keywords{Large Language Model (LLM), Notepad AI, Stylistic Rewrite
}
\received{20 February 2007}
\received[revised]{12 March 2009}
\received[accepted]{5 June 2009}
\maketitle

\section{Introduction}







Natural Language Processing (NLP) is reshaping the world with its rich contextual understanding of text-based systems. Unlike traditional AI, NLP systems primarily rely on textual data. This text contains emotional cues that are often implicit and persistent, making them difficult to suppress without altering the text itself~\cite{kusal2023systematic}. At the same time, style transformation has become a common preprocessing and postprocessing operation in NLP pipelines~\cite{chai2023comparison}. LLMs are now routinely used for style change through prompt engineering text for clarity and other formats by inferring emotion and also act as a framework to quantify how emotional distributions shift before and after style transformation~\cite{zhang2025decoding}. These transformations are typically assumed to preserve semantic meaning while modifying surface-level stylistic features. LMStyle, for example, is a chatbot framework that emphasizes the need for LLMs for text style transfer for intent modeling from large-scale interaction data~\cite{chen2024lmstyle}. However, whether such operations preserve, attenuate, or distort underlying emotional content, less attention has been given in reducing societal harm through stylistic conversation. This research investigates a fundamental yet underexplored question: \textit{Can stylistic rewrite act as a mechanism for emotional harm in social media?} 

In this work, we investigate whether stylistic rewriting can influence the emotional interpretation of text. Through a framework, we show that controlled text transformation can serve as a potential approach to reducing harm in digital communication.

\section{Background}

\subsection{Notepad AI} %

Notepad is a basic text editor that comes built into Microsoft Windows that allows you to create and edit plain text files without formatting. It is lightweight in nature and opens very fast for simple text documents. With the integration of AI writing tools in Notepad, we can shift tone of a text with four distinct styles: formal, casual, inspirational, and humor. \textbf{Formal} tone is structured, precise, and objective, often used for academic writing, business communication, or professional emails. Similarly, \textbf{casual} tone is conversational and relaxed, using simpler language and a friendly rhythm, which works well for social media or informal notes. Furthermore, an \textbf{inspirational} tone focuses on motivation and emotional uplift, often using positive framing and forward-looking language to encourage or energize the reader. Finally, \textbf{humor} tone introduces lightness through wit, playful exaggeration, or irony, making the message more engaging and memorable. Each tone influences not just word choice, but also sentence structure, emotional intensity, and reader perception, so selecting the right one depends on the audience, purpose, and context of communication.

\subsection{Prompt engineering}
Prompt engineering is the process of refining and optimizing inputs to guide LLMs toward generating more accurate, relevant, and high-quality outputs. For example, zero-shot prompt engineering ask the model to perform a task without any prior examples, which can be further enhanced by few show prompting~\cite{marvin2023prompt}.

\subsection{Text Style Transfer}

Text style transfer involves rewriting a sentence to modify its stylistic characteristics while preserving its original semantic meaning. This task requires models to recognize and adjust attributes such as formality, politeness, and sentiment without altering the core content~\cite{liu2023tst}. With the rapid development and strong performance of LLMs across diverse natural language processing tasks, many LLM-based approaches have been proposed for style transfer~\cite{toshevska2025llm}.

\subsection{Emotion Drift}

Emotion Drift in NLP refers to the subtle, dynamic, shift in emotional states throughout a single text or conversation, rather than just assigning one static sentiment in each statement~\cite{jain2023aesthetic}.

\subsection{Related work}
Deep neural networks (DNNs) have been widely used for their ability to automatically learn relevant features from the input text data~\cite{sharma2024review}. The advent of LLMs has remarkably astonished the world due to their emergent capabilities of zero/few-shot learning, in-context learning (ICL), and chain-of-thought, which have revolutionized sentiment and emotion analysis, with the capability of enhanced and accurate sentiment classification in different domains~\cite{luca2024you, zhang2024refashioning}. Due to an LLM's advanced contextual understanding, an LLM can act like a judge due to its reasoning capability and decision-making~\cite{gu2024survey}. Additionally, due to their advanced capability to remember sentiment transitions, these models have demonstrated superior performance. More precisely, these models have shown that they can pick up on both semantic and syntactic contextual relationships~\cite{miah2024multimodal, hung2024novelty}. For instance, Devlin et al.~\cite{devlin2019bert} showed BERT's ability to outperform traditional models in sentiment classification by leveraging its bidirectional context. Similarly, recent studies~\cite{hartmann2023more, chang2024survey} also found that LLMs can surpass traditional sentiment classification models in terms of accuracy and contextual understanding. Furthermore, Mao et al.~\cite{mao2022biases} suggested prompt-based sentiment analysis and emotion detection using pre-trained LLMs. Recently, these models have become increasingly adopted in real-world applications such as advancements in text style transfer using~\cite{toshevska2025llm}. However, there is a lack of significant style change quantification with a standardized approach for stylistic rewrite. In this context, Chen et al.~\cite{chen2025llm} systematically categorize the different types of harms associated with LLMs throughout their lifecycle and propose mitigation strategies, accountability mechanisms, and dynamic auditing frameworks to support more responsible, transparent, and safe integration of LLMs across domains. However, they did not quantize the emotion as a potential mitigation strategy. Soumik et al.~\cite{soumik2025evaluating} experimented on Apple Intelligence to show the privacy-preserving mechanism on Apple devices with stylistic transfer. However, their experiment consisted of a proprietary LLM, and also their developed dataset by prompt engineering is not publicly available. Finally, EmoLLMs, a series of instruction-tuned LLMs supported by a comprehensive multi-task affective dataset and evaluation benchmark, are designed to address both classification and regression tasks in sentiment and emotion analysis in affective analysis~\cite{liu2024emollms}. However, their application is limited to a traditional scope and did not solve the way to reduce harm in a wide-level user context like social media.

\begin{table}[t]
\centering
\small
\renewcommand{\arraystretch}{1.55}
\caption{Our detailed mapping to six basic emotions and their corresponding original emotion labels with qualitative VAD representation. For simplicity, valence, arousal, and dominance are categorized as high (\textcolor{red}H), mid (\textcolor{Orange}M), and low (\textcolor{ForestGreen}L).}
\label{tab:emotion_mapping}
\begin{tabular}{l | p{3.6cm} | p{1.4cm}}
\hline
\textbf{Emotion} & \textbf{Original Emotion Labels} & \textbf{VAD (V,A,D)} \\
\hline

Disgust &
disgust &
(\textcolor{ForestGreen}{L}, \textcolor{Orange}{M}, \textcolor{ForestGreen}{L}) \\
\hline

Anger &
anger, annoyance, disapproval &
(\textcolor{ForestGreen}{L}, \textcolor{red}{H}, \textcolor{Orange}{M}) \\
\hline

Fear &
fear, nervousness &
(\textcolor{ForestGreen}{L}, \textcolor{red}{H}, \textcolor{ForestGreen}{L}) \\
\hline

Sadness &
sadness, disappointment, grief, remorse, embarrassment &
(\textcolor{ForestGreen}{L}, \textcolor{ForestGreen}{L}, \textcolor{ForestGreen}{L}) \\
\hline

Surprise &
surprise, realization, confusion, curiosity &
(\textcolor{Orange}{M}, \textcolor{red}{H}, \textcolor{Orange}{M}) \\
\hline

Happiness &
joy, amusement, excitement, optimism, pride, relief, admiration, approval, gratitude, love, caring, desire &
(\textcolor{red}{H}, \textcolor{red}{H}, \textcolor{red}{H}) \\
\hline

\end{tabular}
\end{table}

\begin{figure}[ht!]
    \centering
    \includegraphics[width=\columnwidth]{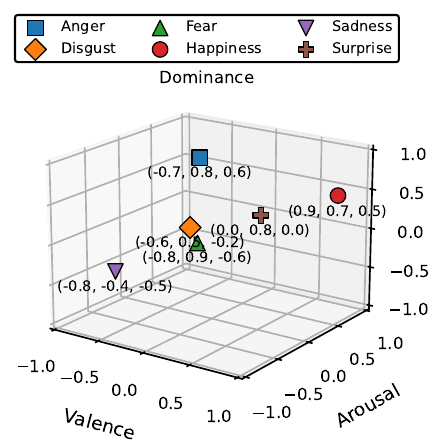}

    \caption{VAD (d=3) analysis of our six basic emotion}
    \label{fig:vad}
\end{figure}

\begin{figure*}[t!]
    \centering
    \includegraphics[width=\textwidth]{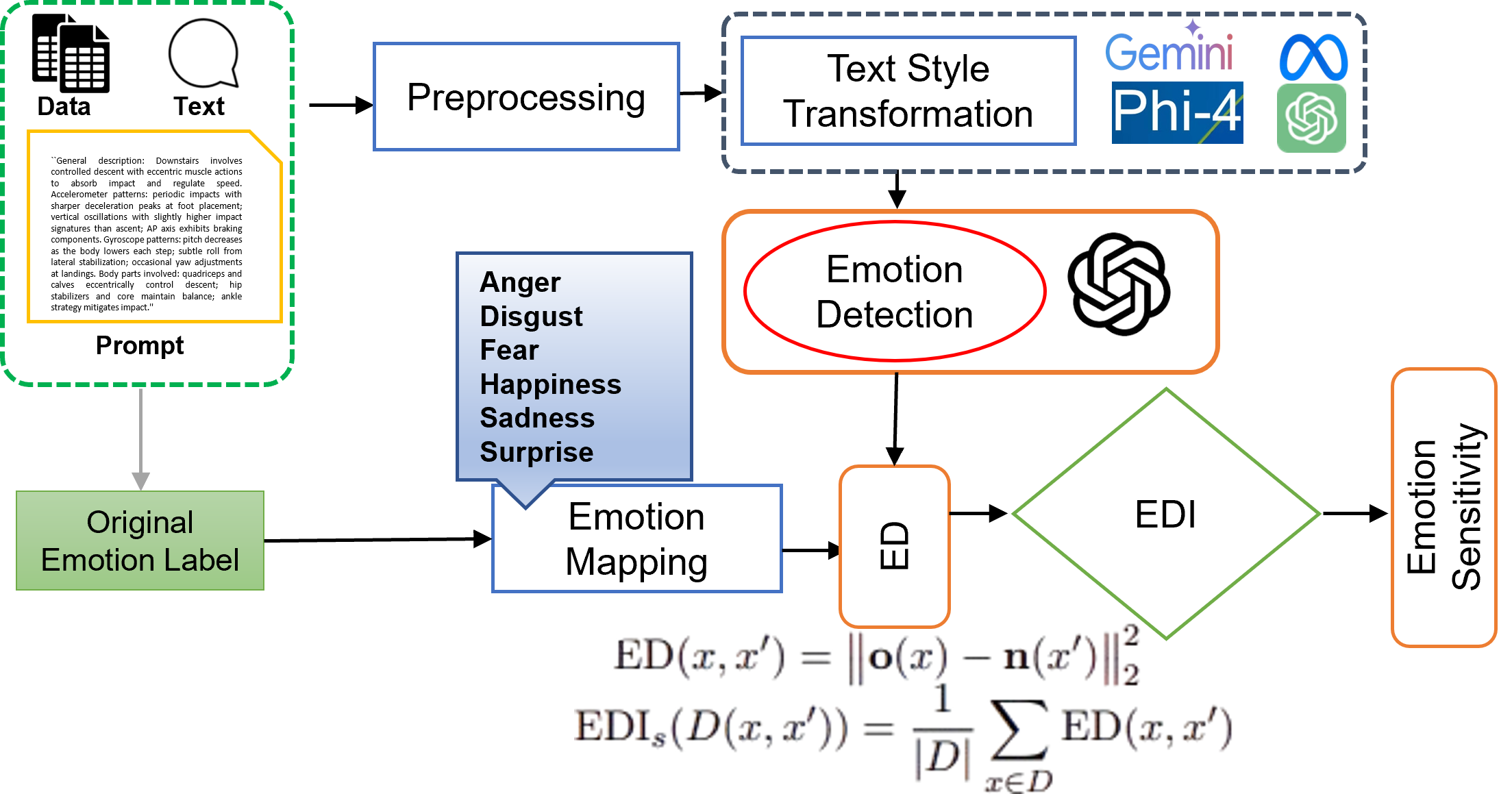}

    \caption{Process of emotion drift analysis to mitigate the harm. We analyze the closest stylistic change through EDI and then analyze the text using LLM and then post the stylistic change content on social media.}
    \label{fig:m}
\end{figure*}

\noindent \textbf{Contribution:}\textit{ We are the first to evaluate and quantify emotion drift induced by stylistic text transformations to prevent societal harm, commonly found in social media. Generally, LLM changes the emotion directly to another emotion; however, there is a gap for quantifying the emotion to the best possible match for a given text. To the best of our knowledge, this study provides the first empirical investigations into stylization-based harm reduction systems, laying the foundation for adaptive rewriting frameworks capable of dynamically neutralizing harmful linguistic patterns to support safer online communication environments.}

\section{System Model}
\textbf{Assumption.} We assume the presence of a filtering mechanism capable of identifying harmful textual content across various posts so that we can moderate the content based on the rewritten comment or post.

\textbf{Attacker Goal.} The attacker aims to spread toxic comments or engage in online harassment in order to cause societal harm to a person or group of people. To achieve this, the attacker posts emotionally charged or harmful content on digital media platforms.

\textbf{Attacker Target.} The primary target is the emotional well-being of the owner of the post. By provoking strong negative reactions, the attacker seeks to create psychological distress, thus creating societal harm to the person who posts and reads the comments.

\textbf{Defender Strategy.} The defender’s objective is to mitigate societal harm by neutralizing the harmful tone of the text. First, the system detects toxic comments or harmful conversations with a filter. Then, the system changes the text through controlled rewriting that reduces emotional intensity while preserving the core meaning of the content. Users are presented only with the moderated version, which is less aggressive and less likely to cause emotional or social disturbance with properly carefully crafted rewritten stylistic text.

\section{Methodology}

Let $x$ denote an original text, and $\mathbf{o}(x)$ defines the associated with ground truth emotion and $x'$ its rewritten version under style
$s \in \{\textsc{formal}, \textsc{casual}, \textsc{inspirational}, \textsc{humor}\}$.
Let $e(x)\in\mathcal{E}$ be the categorical emotion label associated with $x$, where
$\mathcal{E}$ is the vector space of the emotion in n dimensional space where $n = |\mathcal{E}|$.
To quantify how emotion changes induced by stylistic rewriting manifest as reasoning drift, we represent each emotion label as a vector in a continuous affective space. Specifically, we define an embedding function $\mathbf{v}:\mathcal{E}\rightarrow\mathbb{R}^{d}$, where $\mathbf{v}(e)$ is a fixed prototype vector for emotion $e$ , in our case. Valence--Arousal--Dominance (VAD)~\cite{warriner2013norms} circumflex model with dimension $n=3$) \begin{equation} \mathbf{o}(x)=\mathbf{v}(e(x)), \qquad \mathbf{n}(x')=\mathbf{v}(e(x')).\end{equation}
where $\mathbf{v}(.)$ in the VAD domain mapping in the 3$-$dimensional space (Figure~\ref{fig:vad}). We calculate the \textbf{Emotion Drift (ED)} in emotion-space (e.g. VAD domain in figure~\ref{fig:vad}) for a sample as the euclidean magnitude of change between the original and rewritten emotion

\begin{equation}
\mathrm{ED}(x, x')=
 \left\|\mathbf{o}(x)-\mathbf{n}(x')\right\|_{2}^{2}
\label{eq:rdi_vec}
\end{equation}
where value of each emotion the style text emotion found in Figure~\ref{fig:vad} and $\|\cdot\|_{2}^{2}$ denotes the euclidean norm ($\ell_2$).
By construction, $\mathrm{ED}(x, x^{(s)})$ increases when stylistic rewriting produces a larger displacement in emotion space, and it remains near zero when the affective representation is stable.

We propose Emotion Drift Index (EDI), to indicate the degree of emotional change between the original and rewritten versions. For a dataset $D$, we define \textbf{Emotion Drift Index (EDI)} at \textbf{Dataset-Level Aggregation} for a specific style $s$, the  emotion drift index for $D$ for a specific $s$ is computed as
\begin{equation}
\mathrm{EDI}_{s}(D(x, x'))=\frac{1}{|D|}\sum_{x\in D}\mathrm{ED}(x, x').
\label{eq:rdi_dataset}
\end{equation}
where $|D|$ is total size of dataset size with style $s$ with the affective style representation. The higher \textbf{EDI} for a dataset indicates that it is more likely to be affected by a specific style $s$. Figure~\ref{fig:m} illustrates the overall workflow of the proposed framework. In this process, the dataset undergoes style-based rewriting, after which the \textbf{EDI} is applied to measure the magnitude of emotional drift between the original and transformed texts.

\begin{table*}[t]
\centering
\caption{Emotion preservation, change rate, and Emotion Drift Index (EDI) across different writing styles ($s$) of the hateXplain and toxic comment dataset}
\label{tab:emotion_preservation_edi}
\footnotesize
\setlength{\tabcolsep}{3pt}
\renewcommand{\arraystretch}{1.05}

\resizebox{\textwidth}{!}{
\begin{tabular}{l|l|c|c|c|c|c|c}
\hline
\textbf{Dataset} & \textbf{Style} & \textbf{Total} & \textbf{Preserved} & \textbf{Changed} & \textbf{Preserved (\%)} & \textbf{Changed (\%)} & \textbf{EDI$_s$} \\
\hline

\multirow{4}{*}{HateXplain}
 & Formal        & 15383 & 6518 & 8865 & 42.37 & 57.63 & 1.011049 \\
 & Casual        & 15383 & 6166 & 9217 & 40.08 & 59.92 & 0.998109 \\
 & Inspirational & 15383 & 5734 & 9649 & 37.27 & 62.73 & 1.395216 \\
 & Humor         & 15383 & 5636 & 9747 & 36.64 & 63.36 & 1.122511 \\

\hline

\multirow{4}{*}{\shortstack{Toxic\\comment}}
 & Formal        & 15294 & 6523 & 8771 & 42.65 & 57.35 & 1.12025 \\
 & Casual        & 15294 & 5254 & 10040 & 34.35 & 65.65 & 1.244094 \\
 & Inspirational & 15294 & 3605 & 11689 & 23.57 & 76.43 & 1.817063 \\
 & Humor         & 15294 & 4506 & 10788 & 29.46 & 70.54 & 1.326863 \\

\hline
\end{tabular}
}
\end{table*}

\section{Dataset and Tool}

\textbf{Dataset.} This study incorporates widely used harmful and emotion-related text datasets. The \textbf{Toxic Comment Classification Challenge} dataset contains user-generated online comments annotated for multiple forms of toxicity, including toxic, severe toxic, obscene, threat, insult, and identity hate, and is commonly used for modeling abusive language detection~\cite{toxic2018kaggle}. On the other hand, the \textbf{hateXplain} dataset consists of social media posts annotated for hate speech along with target communities and human rationales, enabling fine-grained analysis of hateful and offensive content~\cite{mathew2021hatexplain}.

\textbf{Tools.} As emotion detector, we utilize \textbf{SamLowe/roberta-base-go\_emotions} to convert text to emotion~\cite{lowe2022roberta} trained on Google’s GoEmotions dataset~\cite{demszky2020goemotions}. For consistency, the original 28 GoEmotions labels are mapped into six core emotions: sadness, joy, love, anger, fear, and surprise. To perform stylistic transformation, we apply zero-shot prompting with \textbf{Microsoft Phi-3}~\cite{abdin2024phi} to generate four Notepad AI style variants of each text: \textbf{formal} (structured and professional tone), \textbf{casual} (conversational and relaxed tone), \textbf{inspirational} (motivational and uplifting tone), and \textbf{humor} (light and playful tone). All experiments were conducted on a system with Intel(R) Xeon(R) w5-2465X (3.10 GHz), 32 GB RAM, 4 TB SSD, and 2 × NVIDIA RTX 6000 Ada GPUs, requiring approximately seven days to complete text generation and emotion mapping.

\begin{figure}[t!]
    \centering
    \includegraphics[width=\columnwidth]{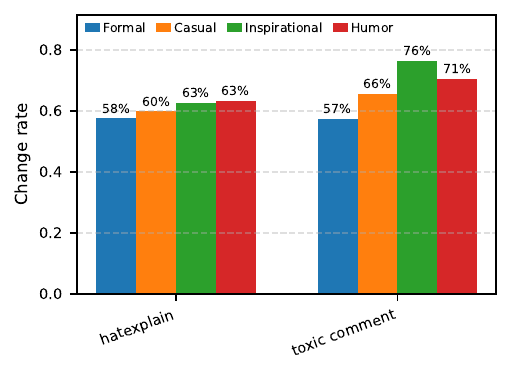}

    \caption{Rate of changes in emotion in two datasets}
    \label{fig:emotion_change_rate}
\end{figure}

\begin{figure*}[htbp]
  \centering
  \begin{tabular}{cccc}
    \includegraphics[width=0.23\linewidth]{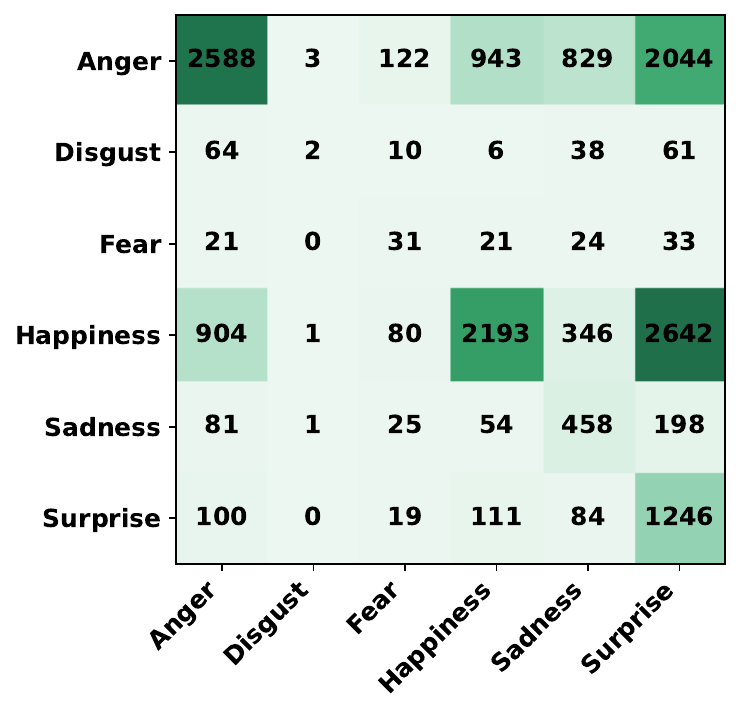} &
    \includegraphics[width=0.23\linewidth]{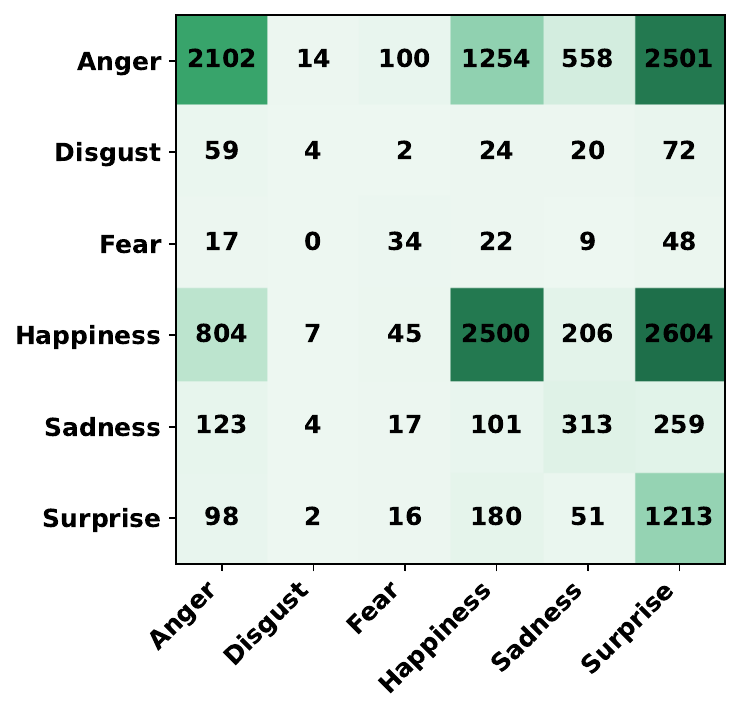} &
    \includegraphics[width=0.23\linewidth]{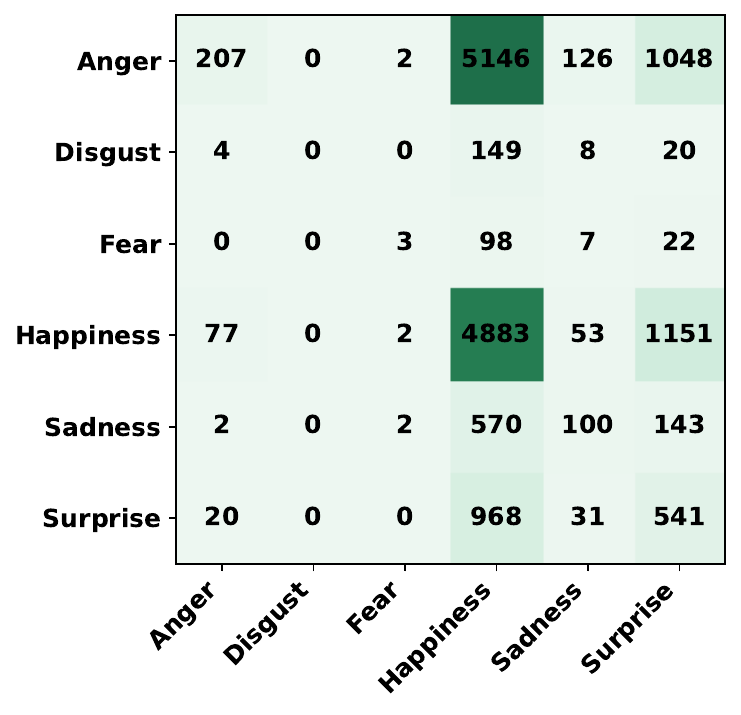} &
    \includegraphics[width=0.23\linewidth]{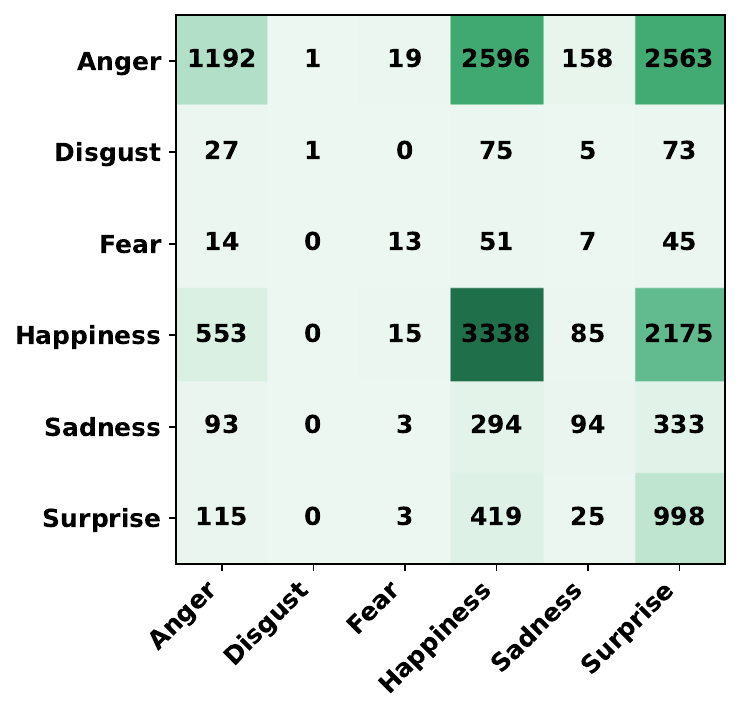} \\
    1(a) Formal & 1(b) Casual & 1(c) Inspirational & 1(d) Humor \\
  \end{tabular}
\begin{tabular}{cccc}
    \includegraphics[width=0.23\linewidth]{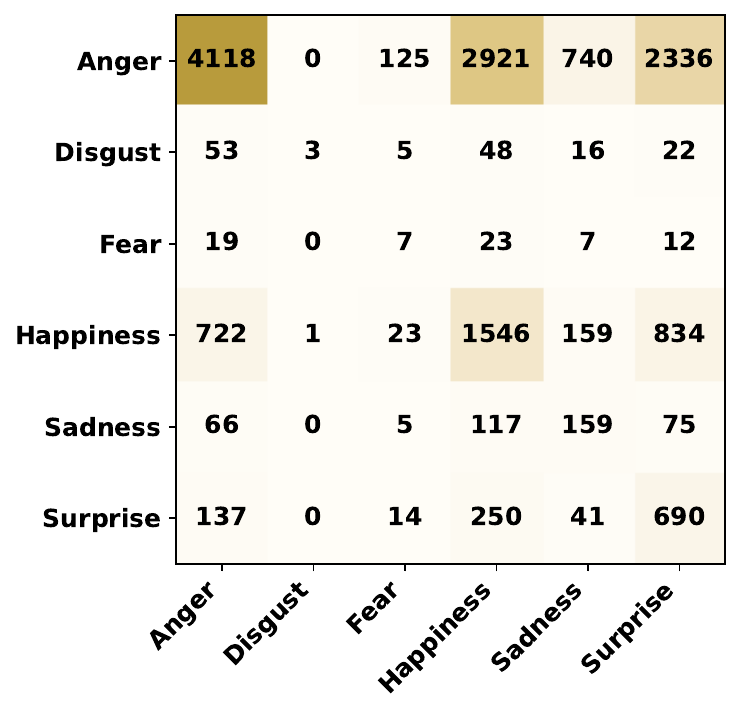} &
    \includegraphics[width=0.23\linewidth]{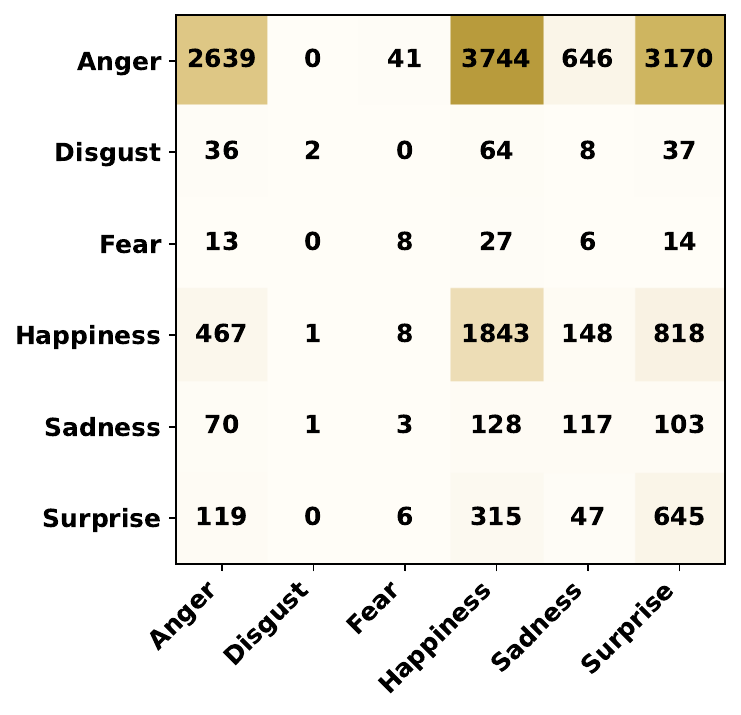} &
    \includegraphics[width=0.23\linewidth]{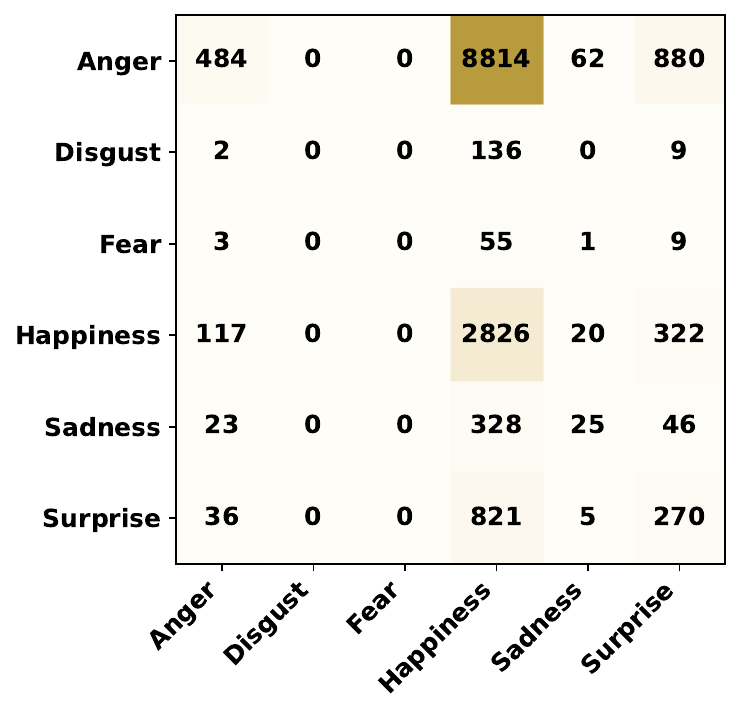} &
    \includegraphics[width=0.23\linewidth]{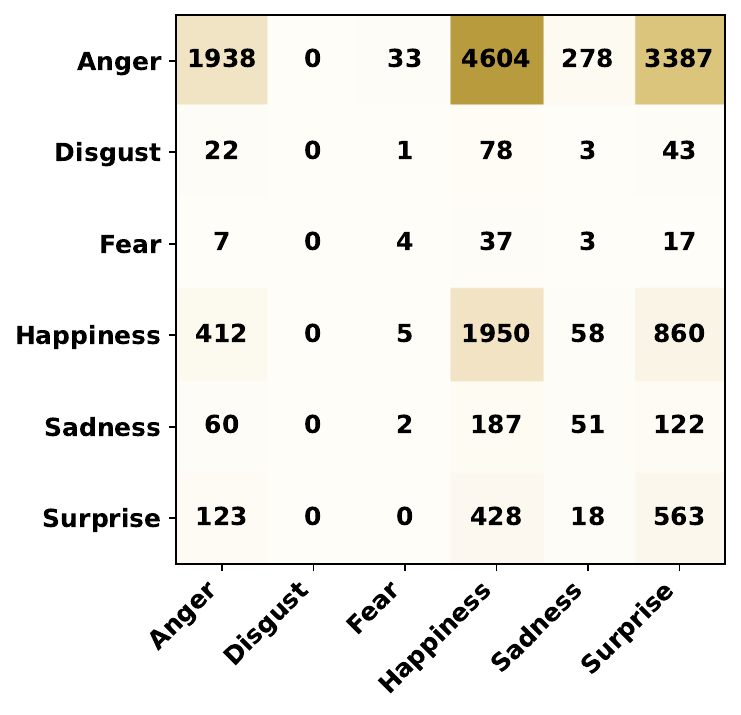} \\
    2(a) Formal & 2(b) Casual & 2(c) Inspirational & 2(d) Humor \\
  \end{tabular}
  
  \caption{Heatmap indicates the most emotion changes of the stylistic rewrite of \textit{ 1. HateXplain and 2. Toxic comment} datasets}

  \label{fig:heatmap}
\end{figure*}

\section{Result Analysis}


\textbf{Pre-processing:} Initially, we analyze the whole dataset and make stylistic transfers and then map our desired emotion of original text and modified texts. Some texts in the datasets may not contain a clearly defined emotion (e.g., neutral). In such cases, we prompt the RoBERTa model to assign the most appropriate emotion label and convert the emotion to the above-mentioned emotion by our defined mapping given in Table~\ref{tab:emotion_mapping} for subsequent analysis.

\textbf{VAD Domain:} We represent discrete emotion categories in a 3D space using fixed VAD prototypes derived from established affective norms. Figure~\ref{fig:vad} illustrates the six emotion categories: anger, disgust, fear, happiness, sadness, and surprise, which are embedded in a three-dimensional VAD space, where each emotion is mapped to a normalized coordinate reflecting its affective profile. Valence captures emotional polarity, arousal reflects intensity, and dominance encodes perceived control in VAD space. The spatial separation among emotions highlights systematic differences in affective structure, with negatively valenced emotions such as anger, fear, and sadness occupying distinct regions from positively valenced emotions such as happiness. This 3D representation provides a quantifying affective positioning that enables us to measure emotion drift and measure Euclidean distance between original and rewritten emotion in VAD space.

\textbf{Emotion Changes Across Data:} Figure~\ref{fig:emotion_change_rate} illustrates how stylistic rewriting affects emotion distributions in the HateXplain and Toxic Comment datasets under four rewriting styles: formal, casual, inspirational, and humor, which visualizes the proportion of samples whose emotion labels changed after rewriting. Each rewriting style introduces measurable emotion transitions, demonstrating that stylistic transformation can significantly alter the perceived emotional content of the text. Figure~\ref{fig:emotion_change_rate} further shows that the Toxic Comment dataset exhibits higher change rates across all styles compared to hateXplain, indicating greater sensitivity to stylistic changes. Additionally, inspirational transformation shows more changes in emotion for both datasets.

Moving forward, we show our overall changes and the correlation provided by the EDI with total changes in Table~\ref{tab:emotion_preservation_edi}. The columns ``total'' have the same values, which indicates that all the text of both datasets are affected by stylistic rewrite. So, the stylistic changes are potentially applicable in a wide variety of cases as well. However, each style has different preserved and changed rates. Our metric, EDI, also relates to the percentage changes; a higher EDI means a higher chance of changing emotions to others. As the LLM changes the text, the semantic meaning of texts is preserved, though the harm is mitigated.

Figure~\ref{fig:heatmap} provides a detailed heatmap of emotion transitions for both datasets. The heatmaps reveal that stylistic rewriting frequently shifts negative emotions e.g. anger and sadness toward more positive or socially acceptable categories like happiness or surprise. This pattern suggests that stylistic transformations particularly inspirational and humorous styles tend to soften emotionally intense language by reframing negative expressions into more positive or neutral affective tones. Overall, the results demonstrate that while stylistic rewriting aims to preserve semantic meaning, it can substantially modify emotional interpretation, highlighting the importance of measuring emotion drift in style-based text transformation systems.

\begin{figure}[h]
    \centering
    \includegraphics[width=\columnwidth]{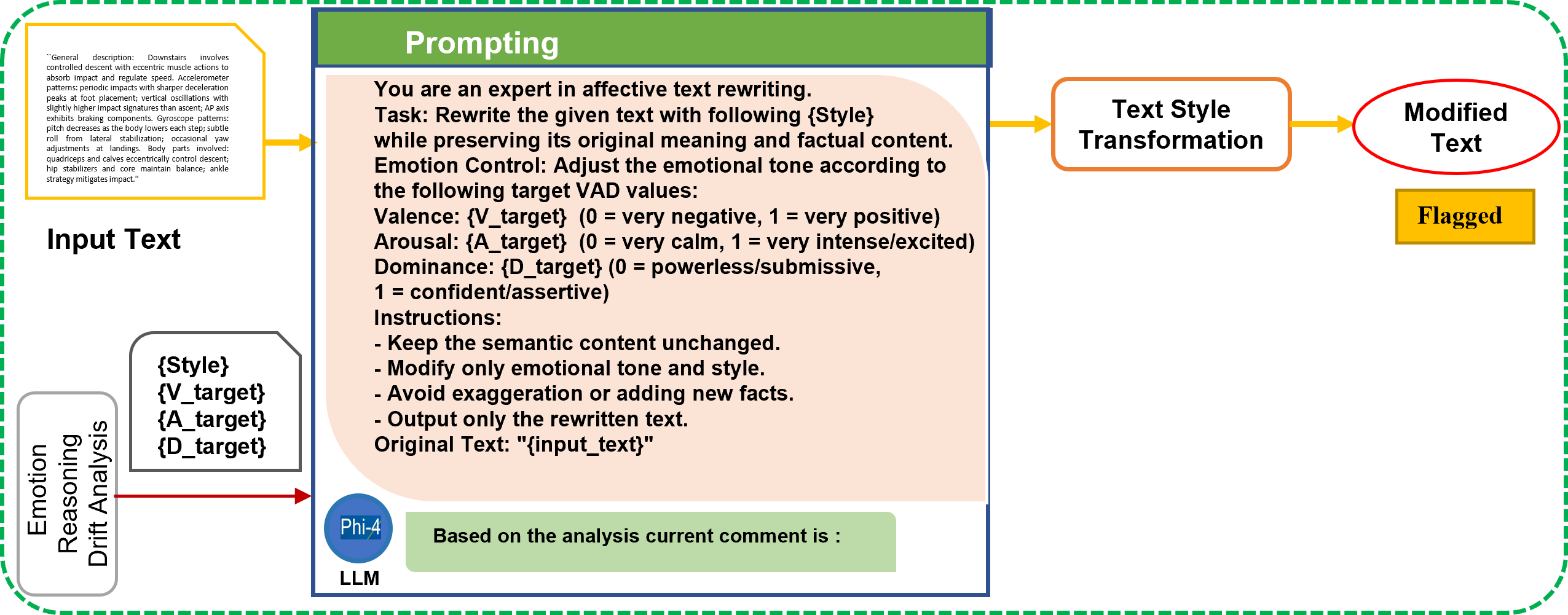}
    \caption{The process of further fine-grained prompt engineering for the style change for converted text}
    \label{fig:rd}
\end{figure}

We further augment this process shown in Figure~\ref{fig:rd} by improving the input text emotion processed using an LLM to generate stylistic rewrites, similar to systems like Notepad AI with a more granular approach shown in Table~\ref{tab:emotion_mapping}. An Emotion Drift Index (EDI) is computed to quantify the variation in emotional representation between the original and rewritten texts. This step enables systematic measurement of potential affective changes caused by stylistic transformation. Based on the measured emotional characteristics, target VAD values are used to generate a controlled prompt shown in Figure~\ref{fig:rd}. The prompt instructs the LLM to adjust the emotional tone of the text while maintaining the original meaning and factual information. Finally, the LLM operates as a decision-making module using few-shot prompting, where predefined examples guide the rewriting process. In this stage, the LLM modifies the emotional tone according to the specified VAD targets. For example, prompt with increasing valence, decreasing arousal, and keeping dominance unchanged can produce a modified text that follows the more desired stylistic and emotional constraints. Finally, the modified text is appended with a notification flag to inform the user of the stylistic conversion.


\section{Conclusion}
This study demonstrates that stylistic rewriting can significantly alter the emotional interpretation of harmful social media comments while preserving their semantic content. The proposed Emotion Drift Index (EDI) provides a quantitative framework for measuring emotional changes introduced by text transformations. Experimental results on the HateXplain and Toxic Comment datasets show that stylistic rewriting frequently shifts negative emotions toward more neutral or positive affective states. These findings suggest that style-based rewriting can serve as a complementary approach to traditional moderation systems by mitigating emotional harm rather than simply removing content.


\section{Limitation}



We use two LLMs in our framework: one for emotion detection (e.g. RoBERTa) and another for text style transformation (e.g. Phi-3). In practice, a single LLM could perform both tasks. However, for our experiments we use two separate models so that the emotion detector remains independent and is not biased by the style changes generated by the same model. Additionally, we do not fine-tune either of the LLMs for the rewriting task, which could improve the generation quality and produce better results, but we intentionally use the base models to evaluate the method under a more conventional and reproducible environment. Furthermore, considering only six basic emotions may have some side effects (e.g. incorrect emotion due to less emotion type); however, for this paper, it serves as a standardized way to demonstrate the step-by-step way to approach the problem of harmful text in social media.

\section{Ethical considerations}
The goal of this research is to explore how controlled stylistic transformation can help reduce emotional harm in online communication. For this research, we do not include harmful or offensive content in this paper.

\bibliographystyle{ACM-Reference-Format}
\bibliography{sample-base}

\appendix
\section*{Language Models}

\section{Microsoft Phi-3}
Microsoft’s Phi-3 model is a compact and high-performing language model designed for enhance reasoning capabilities with fewer parameter counts. The models in the Phi-3 family are trained on carefully crafted web content and high-quality synthetic data, which provide good performance in logical reasoning, mathematics, coding, and structured problem-solving tasks. Additionally, it supports extended context windows and is optimized for deployment on resource-constrained devices, allowing it to run locally on laptops and mobile platforms with reduced latency and improved privacy. Table~\ref{tab:phi_config} provides the detailed configuration for this study.

\begin{table}[h!]
    \centering 
    \caption{Phi-3 Configuration}
    \begin{tabular}{|p{3cm}|p{4cm}|}
        \hline
        \textbf{Configuration} & \textbf{Value} \\
        \hline
        Pretrained Model  & Phi-3-medium-4k-instruct  \\         
        \hline
        Size & medium \\

        \hline
        Context Length    & 4k token\\
        \hline
        Parameters      & 14B \\
        \hline
        
    \end{tabular}
    
    \label{tab:phi_config}
\end{table}

\section{RoBERTa}
The roberta-base-go\_emotions model is a RoBERTa-based transformer fine-tuned on Google’s GoEmotions dataset~\cite{demszky2020goemotions} to perform fine-grained, multi-label emotion classification on English text. Instead of simply predicting positive or negative sentiment, it identifies up to 28 distinct emotional categories including neutral, allowing a single piece of text to express multiple emotions simultaneously. Trained primarily on Reddit comments, the model learns nuanced emotional patterns in language and outputs probability scores for each emotion label. It is particularly useful for social media analysis, mental health research, conversational AI, and customer feedback evaluation, though its performance may vary when applied outside the domain of its training data.
\begin{table}[h!]
    \centering
    \caption{RoBERTa based go emotion Configuration}
    \begin{tabular}{|p{3cm}|p{4cm}|}
        \hline
        \textbf{Configuration} & \textbf{Value} \\
        \hline
        Pretrained Model  & roberta-base \\
        \hline
        Learning Rate     & 2e-5 \\
        \hline
        Weight decay     & 0.01\\
        \hline
        Emotion Level     & 27 $+$ 1 neutral = 28 \\
     \hline
    \end{tabular}
    
    \label{tab:roberta_config}
\end{table}

\section*{Emotion distribution of Datasets}

The original distribution of the original emotion labels in the HateXplain and Toxic Comment datasets in Table~\ref{tab:emotion_distribution}.

\begin{table}[htbp]
\caption{Emotion distribution across HateXplain and Toxic comment datasets}
\label{tab:emotion_distribution}
\centering
\begin{tabular}{ |p{2.4cm} | p{1.9cm} | p{1.9cm} | }
\hline
\textbf{Dataset} & \textbf{Emotion} & \textbf{Count} \\
\hline

\multirow{6}{*}{HateXplain}
 & Anger    & 6529 \\
 & Disgust  & 181  \\
 & Fear     & 130  \\
 & Joy      & 6166 \\
 & Sadness  & 817  \\
 & Surprise & 1560 \\
\hline

\multirow{6}{*}{Toxic Comment}
 & Anger    & 10240 \\
 & Disgust  & 147   \\
 & Fear     & 68    \\
 & Joy      & 3285  \\
 & Sadness  & 422   \\
 & Surprise & 1132  \\
\hline

\end{tabular}
\end{table}

\section*{AI Assistance Disclosure}

Commercial Generative AI tools such as ChatGPT were used for idea generation and language editing during the preparation of this manuscript. Additionally, ChatGPT was also used to explore potential flaws and the clarity of writing. Similarly, we use QuillBot and grammarly for grammar checking and sentence refinement. These tools were used only for writing assistance, and all research design, experiments, analyses, and conclusions were developed by the authors.

\end{document}